\documentclass[a4paper]{article}

\usepackage{INTERSPEECH2019}
\usepackage{amsmath}
\usepackage{epstopdf}
\usepackage{booktabs}
\usepackage{color}
\usepackage{mathtools}
\usepackage{cite}
\usepackage{amssymb}
\usepackage{caption}
\usepackage{subfigure}
\usepackage[ruled]{algorithm2e}

\title{Hard Sample Mining for the Improved Retraining of Automatic Speech Recognition}
\name{Jiabin Xue, Jiqing Han, Tieran Zheng, Jiaxing Guo and Boyong Wu}

\address{
  School of Computer Science and Technology, Harbin Institute of Technology, Harbin, China
  }
\email{\{xuejiabin, jqhan, zhengtieran, guojiaxing, wuboyong\}@hit.edu.cn}
\begin{document}

\maketitle
\begin{abstract}
It is an effective way that improves the performance of the existing Automatic Speech Recognition (ASR) systems by retraining with more and more new training data in the target domain.
Recently, Deep Neural Network (DNN) has become a successful model in the ASR field.
In the training process of the DNN based methods, a back propagation of error between the transcription and the corresponding annotated text is used to update and optimize the parameters.
Thus, the parameters are more influenced by the training samples with a big propagation error than the samples with a small one.
In this paper, we define the samples with significant error as the hard samples and try to improve the performance of the ASR system by adding many of them. Unfortunately, the hard samples are sparse in the training data of the target domain, and manually label them is expensive.
Therefore, we propose a hard samples mining method based on an enhanced deep multiple instance learning, which can find the hard samples from unlabeled training data by using a small subset of the dataset with manual labeling in the target domain.
We applied our method to an End2End ASR task and obtained the best performance.

\end{abstract}

\section{Introduction}
Retraining, a method that trains the existing model with new data in the target domain to improve the performance, is an important topic in Automatic Speech Recognition (ASR) systems \cite{konopka2005retraining, haimi2002automatic}.
The performance of ASR systems is heavily affected by the data scale and domain coverage of training data since they are data-driven.
Thus, the most commonly used retraining method is adding a lot of training data that belongs to the target domain when we try to expend a trained model to be suitable for a target domain.
It is well known that the unlabeled training data can be obtained easily, but it is time-consuming and challenging for the annotation of a large number of unlabeled speech by hand.
Therefore, we wonder: is it possible that just a little data in the whole unlabeled data is manually labeled as the new training data to retrain ASR systems?

Recently, most of ASR systems are based on the Deep Neural Network (DNN)) \cite{sainath2015convolutional, DBLP:conf/icassp/ValtchevOWY96, amodei2016deep}.
In the training process of them, the back propagation (BP) \cite{lecun1990handwritten} method is often used to update the parameters of the systems.
In BP, the systems firstly compute the error between the transcription and the corresponding annotated text.
Then, an error back propagation is used to optimize and update the parameters.
Therefore, it can be found that the parameters are more influenced by the training samples with a big propagation error than the samples with a small one.
We name the former samples as the hard samples and the other as the easy samples.
It represents a different knowledge between the original and the target domain.
Thus, the subset which contains a large number of the hard samples is the one what we want most, and it is useful to improve the ASR retraining via little new training data.

Unfortunately, the hard samples are sparse in the training data of the target domain, and it is expensive to manually label them.
Thus, we try to find them from the transcription in this paper since transcription can fully indicate whether a sentence is misidentified or not.
There are some characters of this task: one sentence is a composition of a series of words, and the correctness of sentence depends on words inside; meanwhile it is easy to mark a whole sentence manually but hard to label each word or words causing the error.
According to above description, we hope to get the label of every word within the sentence by using the sentence-level label only.
So our task can be regard as a weakly supervised learning task.
Since Deep Multiple Instance Learning (DMIL) is one of the  most successful methods of weakly supervised learning \cite{xu2014deep, wu2015deep, kraus2016classifying, papandreou2015modeling, liu2018deep, ilse2018attention}, we select it as a basic method and modify it according to our aim.

\begin{figure}
  \centering
  \subfigure[The distribution of samples in the unlabeled training data]{
    \label{fig:figure1:a}
    \includegraphics[width=1.4in]{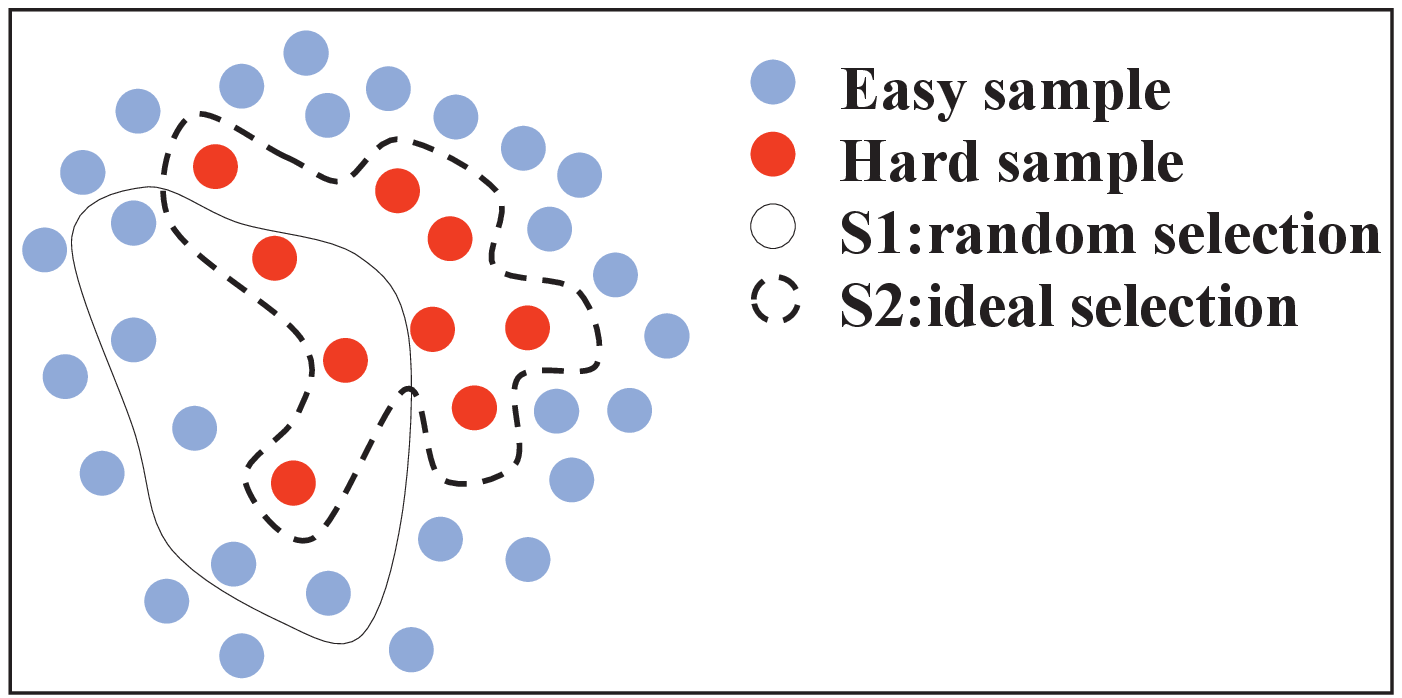}
  }
  \subfigure[The relation between the parameter and character error rate of the target domain]{
    \label{fig:figure1:b}
    \includegraphics[width=1.4in]{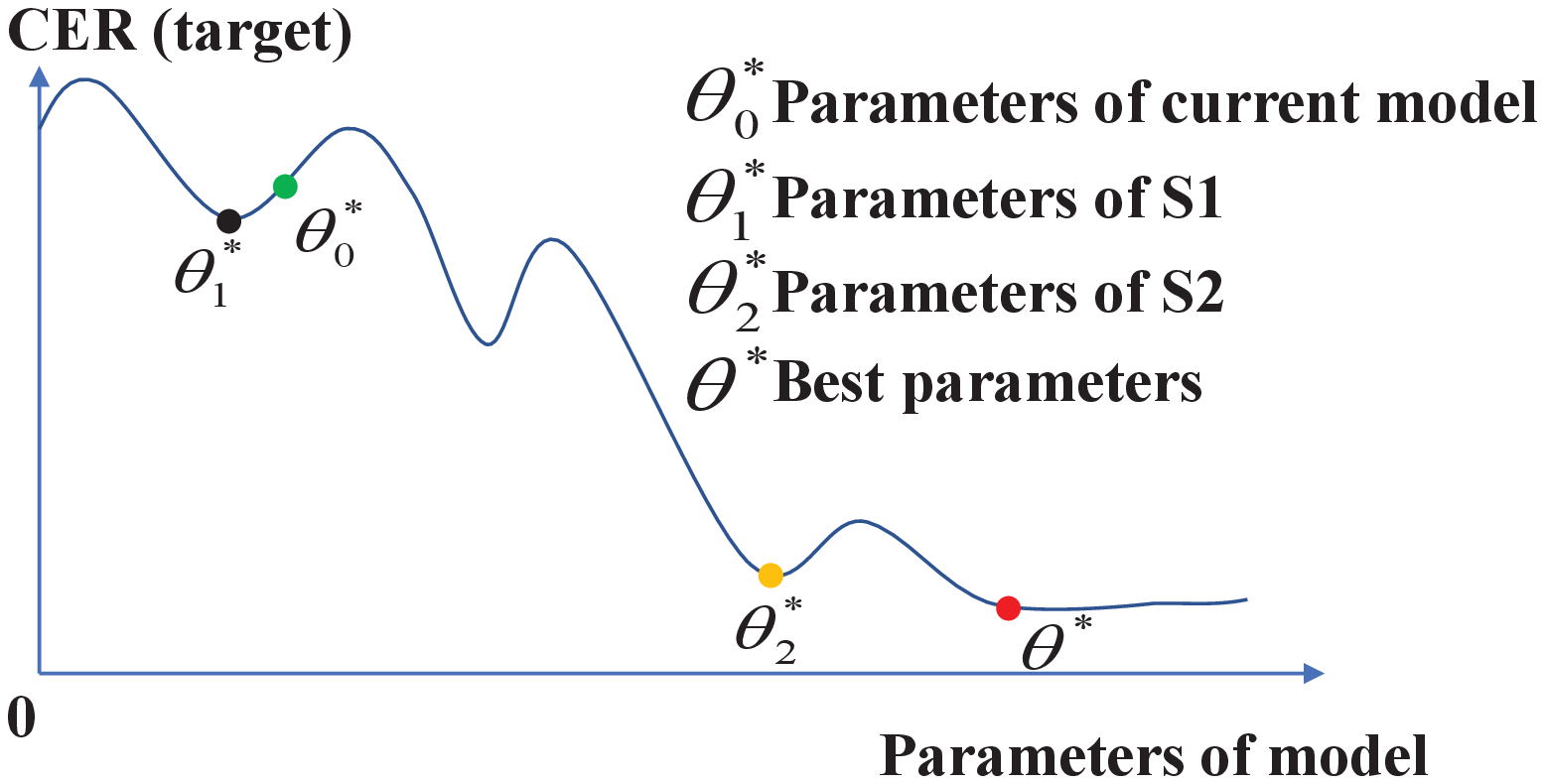}
  }
  \caption{The detailed description of how the hard sample influence on the parameters of the ASR system}
  \label{fig:figure1}
\end{figure}

The contribution of this paper is that we explore a new approach to improve the retraining of ASR by using the hard samples, and we propose three methods, i.e., Sparse-Attention based DMIL, Gated Sparse-Attention based DMIL and Discriminative DMIL, to effectively find the hard samples.

\section{Proposed Method}
\subsection{Retraining Framework Based on Hard Samples}
\subsubsection{Influence of Retraining using Hard Samples }
We elaborate how the hard samples subset influences the performance of the ASR system.
As shown in Figure \ref{fig:figure1}, Figure \ref{fig:figure1}\subref{fig:figure1:a} is assumed as the distribution of the unlabeled data.
When we randomly select a subset from the unlabeled training data, there is a great probability of the $S_{1}$ is selected which contains a small number of the hard samples since the hard samples in the unlabeled data are very sparse.
However, we aim to obtain as many as hard samples as possible, such as the $S_{2}$.
Figure \ref{fig:figure1}\subref{fig:figure1:b} presents the relationship between the parameters and character error rate (CER) of the target domain, where $\theta_{0}^{*}$ is the parameters of current model and $\theta^{*}$ is the best parameters in the target domain.
If we select the subset $S_{1}$ to train the current system, we will find $\theta_{1}^{*}$ near to $\theta_{0}^{*}$ fits well, but it is not suitable for the whole target domain.
On the contrary, if the subset $S_{2}$ which includes lots of hard samples is selected to train the current model, we can find $\theta_{2}^{*}$ fits well and it is very close to the $\theta^{*}$ suitable for the whole target domain.
The reason is that there is a great difference in distribution between the hard samples in the target domain and the samples in the original training data, but for the easy samples, there is not.
In summary, the hard samples overwhelmingly promote the model retraining process, thus we attempt to improve the ASR system by using subset full of the hard samples like $S_{2}$.

We define the hard samples subset as follows:
\begin{equation}
\begin{aligned}
\label{eq1}
H &= \{x_{i} \mid x_{i} \notin A \cup x_{i} \in B\} , \forall i \in \mathbb{R},
\end{aligned}
\end{equation}
where, $H$ is the hard samples subset, $A \in \mathbf{D}_{A}$ is the training data of current domain, ${B \in \mathbf{D}_{B}}$ is the training data of target domain and $x_{i}$ is the $ith$ data sample.

\subsubsection{Hard Samples Mining based ASR Retraining Framework}
\begin{figure}
\centering
\includegraphics[width=3.2in]{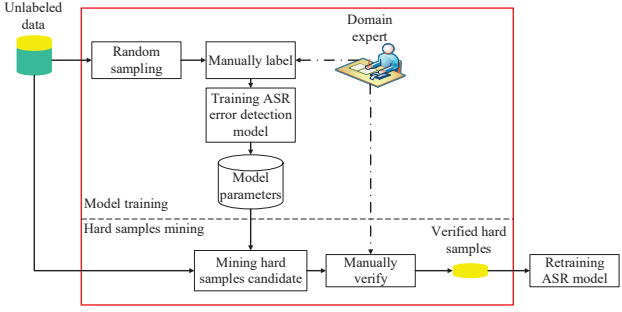}
\caption{The process of the proposed retraining framework}
\label{fig:side:a}
\end{figure}

Based on the problem described above, we attempt to improve the retraining of ASR systems by using the hard samples subset and propose a novel framework to retrain the ASR system.
The structure of the framework is shown in Figure \ref{fig:side:a}, and it contains three parts as follows.
In the first part, we collect a lot of unlabeled target domain data. In the second part, we mine the hard samples subset from the unlabeled data by using a ASR error detection model and then manually label them.
In the third part, the ASR systems parameters are updated by using the hard samples.

In this process, mining hard samples from unlabeled target domain data is an extremely critical step.
Thus, we introduce the method about how to mine hard samples from the unlabeled training.
As shown in the red box of Figure \ref{fig:side:a}, this method can be divided into two parts, including training ASR error detection model and mining the hard samples from unlabeled data by using the trained error detection model.
First, we select a subset according to a stationary sampling interval from the unlabeled data which has been sorted before by the sentence length.
Then, the selected subset is input into the original ASR system and it will output the corresponding transcription.
The domain experts label the transcription by comparing the received transcription and corresponding sentence: if they are identical, mark $1$; otherwise mark $0$.
In this way, all transcriptions are labeled with $0$ or $1$.
Next, the labeled transcription is used to train the ASR error detection model.
Finally, the trained ASR error detection model is used to mine the candidate hard samples from the unlabeled data, and the candidate hard samples will be further manually labeled to determine the final results.

\subsection{Hard Samples Mining by Using DMIL}
\subsubsection{Problem Formulation}
In this paper, our aim is to find the hard samples from lots of unlabeled target domain data. Because it can be used to improve the retraining of ASR systems, we propose to use transcription to achieve our aim.

It is well known that the sentence consists of a series of words and there are some context relationship between them.
We often use a fixed-length context window to divide the transcription into multiple context text blocks to consider the context between them.
Thus, we determine the class of the transcription based on the classification result of each text block.
Further, we can write the description by the form as follows:

\begin{equation}
\label{eq1}
{X_{i} = \{\mathbf{x}_{i}^{0}, \mathbf{x}_{i}^{1}, ..., \mathbf{x}_{i}^{N-1}\}},
\end{equation}
\begin{equation}
\label{eq2}
\hat{Y_{i}}=\left \{
              \begin{aligned}
                &0,            & &\mathbf{if} \quad {\sum_{k=0}^{N-1}}y_{i}^{k}=0,  \\
                &1,  & &\mathbf{otherwise},   \\
              \end{aligned}
        \right.
\end{equation}
where $X_{i}$ is the $ith$ sentence generated by the ASR system.
$N$ is the number of words in the $X_{i}$.
$\mathbf{x}_{i}^{j} \in \mathbb{R}^{T \times D}$ represents the feature vector of the $jth$ word in the $ith$ sentence.
$D$ is the dimension of the word feature and $T$ is the width of the context window.
$y_{i}^{k} \in \{0,1\}$ is the word label of $\mathbf{x}_{i}^{k}$ and $Y_{i} \in \{0,1\}$ is the sentence label of $X_{i}$.
And we define the $\mathbf{x}_{i}^{k}$ as key word if $y_{i}^{k} = 1$.
We hope that the ASR error detection model learns $y_{i}^{j}$ while learning $Y_{i}$ when we input the sample pair $\{X_{i},Y_{i}\}$ into it.


\subsubsection{Attention based DMIL}
We use the Attention based DMIL model \cite{ilse2018attention} as the baseline in this paper, which includes word level embedding network, pooling, and classifier network.
In this structure, the word level embedding network is responsible for extracting the feature of the text block.
The pooling is used to compress multiple word features into one sentence feature.
And the classifier network predicts the class of sentence by using this feature.

When we input a sample pair $\{X_{i},Y_{i}\}$, the whole process is as follows:
\begin{equation}
\label{eq3}
{\mathbf{h}_{i}^{j} = \Phi(\mathbf{x}_{i}^{j})},
\end{equation}
\begin{equation}
\label{eq4}
{\mathbf{e}_{i}^{j} = \mathbf{W}^{T}(\tanh(\mathbf{V}\mathbf{h}_{i}^{j^{T}}))},
\end{equation}
\begin{equation}
\label{eq5}
{\mathbf{a}_{i} = softmax(\mathbf{e}_{i})},
\end{equation}
\begin{equation}
\label{eq6}
{\mathbf{H}_{i} = \sum_{j=0}^{N-1}\mathbf{a}_{i}^{j}\mathbf{h}_{i}^{j}},
\end{equation}
\begin{equation}
\label{eq7}
{\tilde{Y_{i}} = \Psi(\mathbf{H}_{i})},
\end{equation}
where $\Phi(\cdot)$ is the word level embedding network, $\mathbf{h}_{i}^{j} \in \mathbb{R}^{D}$ is the D-dimensional feature vector of the word $\mathbf{x}_{i}^{j}$.
$\mathbf{W}$ and $\mathbf{V}$ are the weight matrices of the attention mechanism.
$\mathbf{e}_{i}^{j}$ and $a_{i}^{j}$ are the score and the attention value of the $\mathbf{h}_{i}^{j}$ respectively.

\subsubsection{Hard Samples Mining using Sparse-Attention}
For the attention based on DMIL, with the improvement of the front-end ASR system, the number of errors in the transcription is decreasing, and the similarity of the right and error transcription is increasing.
In this case, the traditional Attention based DMIL cannot efficiently find error in the transcription.

In the traditional attention mechanism, we calculate the score of each word through a fully-connected network, and then normalize the score by using the softmax transformation function:
\begin{equation}
\label{eq8}
{softmax(\mathbf{e}_{i}) = \frac{\exp(\mathbf{e}_{i}^{j})}{\sum_{k}\exp(\mathbf{e}_{i}^{k})}}.
\end{equation}

However, the softmax transformation function projects the input into a normalized output vector of the specified dimension, while each dimension of the obtained output is greater than zero; this is wasteful.
Moreover, the attention value of each word may be close to the average value, which results the situation that the key word cannot be found.

With this in mind, we propose an improved DMIL from the aspect of attention mechanism, i.e., Sparse-Attention based DMIL. We replace the \eqref{eq5} by the sparsemax translation function, which is as follows:
\begin{equation}
\label{eq9}
{sparsemax(\mathbf{e}_{i}) = \arg\min_{\bm{\alpha} \in \Delta^{j}}\|\bm{\alpha} - \mathbf{e}_{i}\|},
\end{equation}
where $\Delta^{j} \coloneqq \{\bm{\alpha} \in \mathbb{R}^{j} | \bm{\alpha} \geq 0, \sum_{j}\bm{\alpha}_{j}=1\}$.
In other words, it is the Euclidean projection of the scores $\mathbf{e}_{i}$ onto the probability simplex.
These projections tend to hit the boundary of the simplex and yield a sparse probability distribution.
This allows the classifier to attend only to a few words in the sentence and assign zero probability mass to all other words.
It has been shown that the asymptotic cost of sparsemax and softmax is the same and the gradient back-propagation of sparsemax is faster than that of softmax which takes sublinear time.

\subsubsection{Hard Samples Mining using Gated Sparse-Attention}
To further enhance the ability of Sparse-attention based DMIL, we notice that it is difficult to learn complex relations efficiently by using the $tanh(\mathbf{z})$ in \eqref{eq4}.
Our concern follows from the fact that $tanh(\mathbf{z})$ is approximately linear for $\mathbf{z} \in [-1, 1]$, which
probably limits the final expressiveness of the learned relations among words.
Thus, we propose to replace the \eqref{eq4} and \eqref{eq5} by the Gated Sparse Attention mechanism, which additionally uses the gating mechanism \cite{DBLP:conf/icml/DauphinFAG17} together with $tanh(\cdot)$ that yields:
\begin{equation}
\label{eq10}
{\mathbf{e}_{i}^{j} = \mathbf{W}^{T}(\tanh(\mathbf{V}\mathbf{h}_{i}^{j^{T}}))\odot sig(\mathbf{U}\mathbf{h}_{i}^{j^{T}})},
\end{equation}
\begin{equation}
\label{eq11}
{\mathbf{a}_{i} = sparsemax(\mathbf{e}_{i})},
\end{equation}
where $\mathbf{U}$ is the weight matrix, $\odot$ is an element-wise multiplication and $sig(\cdot)$ is the sigmoid activation function.
The gating mechanism probably removes the problem in $\tanh(\cdot)$ by introducing a learnable non-linearity.

\subsubsection{Hard Samples Mining using Discriminative Embedding}
In traditional DMIL, we usually use the softmax function as the activation function for the last layer of the classifier network.
However, there are two disadvantages, which lead it cannot find the key word effectively.
On the one hand, as the similarity of the training data increases, the classifier network structure will become complex.
Meanwhile, the gradient of the attention will be very small, even encounter the problem of vanishing gradient.
On the other hand, many studies have shown that the softmax function is unable to effectively guide the training of the embedding network and it is difficult to find key word effectively \cite{DBLP:conf/icml/LiuWYY16, DBLP:conf/cvpr/LiuWYLRS17}.
Thus, the performance of the traditional DMIL is limited.

In order to solve the problem mentioned above, we also propose the SVM-based DMIL, which uses a two-stage training strategy to train the word embedding network and the classifier network separately.
In the first training stage, we optimize the SVM and embedding network jointly to make the obtained embedding more discriminating.

The original SVM is used to solve the binary classification problems.
Given the training sample pair ${\{\mathbf{H}_{i},Y_{i}\}}$, ${i \in [1,2, ..., N]}$, the SVM optimizes the following constraint problems:
\begin{equation}
\label{eq12}
{\min_{\mathbf{w},\xi_{n}}\frac{1}{2}\mathbf{w}^{T}\mathbf{w} + c\sum_{n=1}^{N}\xi_{n}},
\end{equation}
\begin{equation}
\label{eq13}
{s.t. \quad \mathbf{w}^{T}\mathbf{w}\mathbf{H}_{n}Y_{n}\geq 1 - \xi_{n} \quad \forall n},
\end{equation}
\begin{equation}
\label{eq14}
{\xi_{n} \geq  0 \quad \forall n},
\end{equation}
where the $\xi_{n}$ is the slack variables, which are used to penalize the misclassified samples.
The $c$ is the penalty factor, which controls the penalty size for misclassified samples.
At the same time, the selection of $c$ will greatly affect the training speed of the neural network.

Then, we can convert the above optimization problem to an unconstrained optimization problem as follows:

\begin{equation}
\label{eq15}
{\min_{\mathbf{w}}\frac{1}{2}\mathbf{w}^{T}\mathbf{w} + c\sum_{n=1}^{N}\max(1 - \mathbf{w}^{T}\mathbf{H}_{n}Y_{n}, 0)}.
\end{equation}

Further, we can convert \eqref{eq15} into a neural network objective function.
\begin{equation}
\label{eq16}
{\arg\min_{f}\frac{1}{2}\mathbf{w}^{T}\mathbf{w} + c\sum_{n=1}^{N}g(1 - \mathbf{w}^{T}\Psi(\mathbf{H}_{n})Y_{n})},
\end{equation}
where $g(\mathbf{z})=\frac{1}{\beta}log(1 + exp({\beta}\mathbf{z}))$ is the generalized logistic loss function, which is a smoothed approximation of the hinge loss function $[\mathbf{z}]_{+} = \max(\mathbf{z}, 0)$, $\beta$ is a sharpness parameter \cite{DBLP:conf/cvpr/MignonJ12}.
In this paper, we use the fixed $\beta = 1$ to reduce the number of hyperparameters.

\begin{table*}[!htbp]
\centering
\caption{Experimental Results for ASR Error Detection}
\begin{tabular}{ c | c | c | c | c | c | c | c }
\hline
Model Name         & Type of Model                       & SI   & $ACC (sentence)(\%)$  & $P$     & $R$    & $F$ &$ACC(word)(\%)$\\ \hline
$baseline$            & TextCNN                          & $-$  & 82.6 &  $-$ &  $-$ & $-$  & $-$ \\

$A\_3\_1$      & Attention based DMIL                    & $3$  & 82.8  &  $0.53$ &  $0.72$ & $0.61$  & $93.0$\\

$A\_5\_1$      & Attention based DMIL                    & $5$  & 81.9 &  $-$ &  $-$ & $-$  & $-$ \\

$A\_7\_1$      & Attention based DMIL                    & $7$  & 81.7 &  $-$ &  $-$ & $-$  & $-$ \\

$SA\_3\_1$     & Sparse-Attention based DMIL             & $3$  & 82.5  &  $0.94$ &  $0.37$ & $0.53$  & $91.3$\\

$GSA\_3\_1$    & Gated-Sparse-Attention-based DMIL       & $3$  & 82.6 &  $0.87$ &  $0.49$ & $0.63$  & $91.2$\\

$ADT\_3\_1$    & Attention-based DMIL + DT               & $3$  & 83.0  &  $0.64$ &  $0.68$ & $0.66$  & $94.6$ \\

$SADT\_3\_1$   & Sparse-Attention-based DMIL + DT        & $3$  & 82.7 &  $0.88$ &  $0.53$ & $0.66$  & $93.1$ \\

$GSADT\_3\_1$  & Gated-Sparse-Attention-based DMIL + DT  & $3$  & 82.9  &  $0.79$ &  $0.61$ & $0.69$  & $95.7$
\\\hline
\end{tabular}
\label{Table1}
\end{table*}

\section{Experimental Details}

\subsection{Experiment Settings}
In this section, we shall introduce the datasets, the structure of models, and the training strategies which are used in the experiments on this paper.

Our experiments are conducted on the 300 hours Switchboard English conversational telephone corpus \cite{godfrey1992switchboard} and the 2000 hours Fisher corpus, which are the most studied ASR benchmark today \cite{xiong2017microsoft, vesely2013sequence, povey2016purely, medennikov2016improving}.

We have trained a Listen Attend and Spell (LAS) \cite{chan2016listen} model with ESPnet \cite{DBLP:conf/interspeech/WatanabeHKHNUSH18} tools \footnotemark[1] on SwitchBoard corpus.
\footnotetext[1]{The code of ESPnet is available at
https://github.com/espnet/espnet }
We select 100h training data from the Fisher corpus to train the ASR error detection model and then use it to mine the hard samples from the rest of this corpus.

For various of the DMIL models, we design the word level embedding network as a Convolutional Neural Network (CNN) for extract the word level feature, the pooling as two layers fully-connected network, and use three layers fully-connection neural network with Rectified Linear Unit (ReLu) activation function as the classifier network.

The training strategies in this paper are as follows.
First, the weights for all layers are uniformly initialized to lie between -0.05 and 0.05.
Then networks are trained using Adam \cite{DBLP:journals/corr/KingmaB14} with a learning rate of 0.001.
The learning rate is halved whenever the held-out loss does not decrease by at least 10\%.
Finally, we clip the gradients to lie in $(-10, 10)$ to stabilize training.

\subsection{Results of Sentence Level ASR Error Detection}
For evaluating the sentence classification ability of our methods, we choose the text Convolutional Neural Network classification (TextCNN), which is widely used in the field of sentence classification, as the baseline model \cite{DBLP:conf/emnlp/Kim14} \footnotemark[2].
\footnotetext[2]{The code of TextCNN is available at https://github.com/dennybritz/
cnn-text-classification-tf }

We shall analyze the results of the sentence level ASR error detection model.
All the model structures and experimental results are described in Table \ref{Table1}.
Where $SI$ is the width of the context window, and the $ACC (sentence)$ is the accuracy of the sentence detection.

First, we explore the performance of Attention based DMIL on the sentence level ASR error detection model.
By comparing the accuracy of the $baseline$ and $A\_3\_1$, we found that it has similar performance with the $baseline$ model in this task.

Then, we try to explore the influence of the width of the context window $ST$.
We compare the models with various $ST$, which include $A\_3\_1$, $A\_5\_1$ and $A\_7\_1$.
We have found that the best performance of Attention based DMILs is $82.8\%$ with $A\_3\_1$ and the performance of them is decreased with $ST$ increases.
Thus, in the next experiments, we use $ST=3$.

Finally, we explore the performance of the improved methods.
We find the Sparse-Attention and the Gated Sparse-Attention are unable to improve the performance of Attention based DMIL by comparing $A\_3\_1$, $SA\_3\_1$ with $GSA\_3\_1$.
We also compare $A\_3\_1$ with $ADT\_3\_1$ and compare $GSA\_3\_1$ with $GSADT\_3\_1$ at the same time.
The discriminative training (DT) strategy can significantly improve the performance of DMIL, and the best performance of these models is $83.0\%$.
\subsection{Results of Word Level ASR Error Detection}
For evaluate the word classification ability of our methods, we choose the $A\_3\_1$ as the baseline model.
We analyze the results of the word level ASR error detection model.
All the model structures and experimental results are described in Table \ref{Table1}.
Where the $P$,$R$, $F$ and $ACC(word)$ are the Precision, Recall, F1 Score and Accuracy.

First, we explore the performance of Attention based DMIL on the word level ASR error detection model.
From the second row in Table \ref{Table1}, we can find that the $P$ is close to $0.5$, thus it cannot find the key words efficiently. The reason is that the shortcoming of the attention mechanism with the softmax.

Then, we explore the performance of the Sparse-Attention DMIL model.
By observing the recorded value of the fifth row in Table \ref{Table1}, we found that it can find a small number of key words with high accuracy and cannot find most of the key instances, and the F1 score is lower than the baseline model.

Next, we shall explore the influence of the gating mechanism.
By comparing $SA\_3\_1$ and $GSA\_3\_1$, we can find that it can improve the performance of Sparse-Attention based DMIL by helping it to find more key words, which proves our previous assumption that the gating mechanism probably removes the problem in $tanh(\cdot)$ by introducing a learnable non-linearity.

Finally, we try to explore the influence of the DT strategy.
We compare $A\_3\_1$ and $ADT\_3\_1$ and find that the DT strategy can improve the precision of the Attention based DMIL.
And then, we compare $A\_3\_1$ with $ADT\_3\_1$ and compare $GSA\_3\_1$ with $GSADT\_3\_1$ at the same time.
We get the same conclusion as the first comparison.

\begin{table}[!htbp]
\centering
\caption{Performance of the retrained ASR via hard samples}
\begin{tabular}{ c | c | c }
\hline
Dataset                      & Size(h)    & WER (\%)  \\ \hline

SwitchBoard                  &  300       & $22.31$    \\

SwitchBoard + Easy samples  &  800      & $20.3$ \\

SwitchBoard + Hard samples  &  800      & $\textbf{17.2}$

\\\hline
\end{tabular}
\label{Table3}
\end{table}
\subsection{Results of Retraining  ASR via Hard Samples}
We show the results of the retrained ASR system via the hard samples with different datasets in Table \ref{Table3}.
We can see that the model with a 500h hard samples dataset gets the best performance in the switchboard test set.
\section{Conclusions}
In this paper, we first propose an improved retraining framework for ASR by using the hard samples.
And then, we propose a novel method, which is based on Attention based DMIL, for mining the hard samples from unlabeled data.
This method is able to find locations of errors while determining the class of transcription.
Furthermore, we propose three enhanced methods from the attention mechanism and training strategy respectively, i.e., Sparse-Attention based DMIL, Gated Sparse-Attention based DMIL and DDMIL.
We verified our proposed methods on the SwitchBoard corpus and Fisher corpus.
From the experiment results, we can see that compared with the traditional training method, the retrained model via hard samples gets great improved performance.

\section{Acknowledgements}
This research was supported by the National Key Research and Development Plan of China under Grant 2017YFB1002102 and National Natural Science Foundation of China under Grant U1736210

\clearpage
\bibliographystyle{IEEEtran}
\bibliography{mybib}

\begin{thebibliography}{10}
\providecommand{\url}[1]{#1}
\csname url@samestyle\endcsname
\providecommand{\newblock}{\relax}
\providecommand{\bibinfo}[2]{#2}
\providecommand{\BIBentrySTDinterwordspacing}{\spaceskip=0pt\relax}
\providecommand{\BIBentryALTinterwordstretchfactor}{4}
\providecommand{\BIBentryALTinterwordspacing}{\spaceskip=\fontdimen2\font plus
\BIBentryALTinterwordstretchfactor\fontdimen3\font minus
  \fontdimen4\font\relax}
\providecommand{\BIBforeignlanguage}[2]{{%
\expandafter\ifx\csname l@#1\endcsname\relax
\typeout{** WARNING: IEEEtran.bst: No hyphenation pattern has been}%
\typeout{** loaded for the language `#1'. Using the pattern for}%
\typeout{** the default language instead.}%
\else
\language=\csname l@#1\endcsname
\fi
#2}}
\providecommand{\BIBdecl}{\relax}
\BIBdecl

\bibitem{konopka2005retraining}
C.~Konopka and L.~C. Almstrand, ``Retraining and updating speech models for
  speech recognition,'' Sep.~6 2005, {US Patent 6,941,264}.

\bibitem{haimi2002automatic}
R.~Haimi-Cohen, ``Automatic retraining of a speech recognizer while using
  reliable transcripts,'' Apr.~16 2002, {US Patent 6,374,221}.

\bibitem{sainath2015convolutional}
T.~N. Sainath, O.~Vinyals, A.~W. Senior, and H.~Sak, ``{Convolutional, Long
  Short-Term Memory, fully connected Deep Neural Networks},'' in \emph{{IEEE}
  International Conference on Acoustics, Speech and Signal Processing,
  {ICASSP}}, 2015, pp. 4580--4584.

\bibitem{DBLP:conf/icassp/ValtchevOWY96}
V.~Valtchev, J.~Odell, P.~C. Woodland, and S.~J. Young, ``Lattice-based
  discriminative training for large vocabulary speech recognition,'' in
  \emph{{IEEE} International Conference on Acoustics, Speech, and Signal
  Processing Conference Proceedings, {ICASSP}}, 1996, pp. 605--608.

\bibitem{amodei2016deep}
D.~Amodei, S.~Ananthanarayanan, R.~Anubhai, J.~Bai, E.~Battenberg, C.~Case,
  J.~Casper, B.~Catanzaro, Q.~Cheng, G.~Chen \emph{et~al.}, ``Deep speech 2:
  End-to-end speech recognition in english and mandarin,'' in
  \emph{International conference on machine learning}, 2016, pp. 173--182.

\bibitem{lecun1990handwritten}
Y.~LeCun, B.~E. Boser, J.~S. Denker, D.~Henderson, R.~E. Howard, W.~E. Hubbard,
  and L.~D. Jackel, ``Handwritten digit recognition with a back-propagation
  network,'' in \emph{Advances in neural information processing systems}, 1990,
  pp. 396--404.

\bibitem{xu2014deep}
Y.~Xu, T.~Mo, Q.~Feng, P.~Zhong, M.~Lai, and E.~I. Chang, ``Deep learning of
  feature representation with multiple instance learning for medical image
  analysis,'' in \emph{{IEEE} International Conference on Acoustics, Speech and
  Signal Processing, {ICASSP}}, 2014, pp. 1626--1630.

\bibitem{wu2015deep}
J.~Wu, Y.~Yu, C.~Huang, and K.~Yu, ``Deep multiple instance learning for image
  classification and auto-annotation,'' in \emph{{IEEE} Conference on Computer
  Vision and Pattern Recognition, {CVPR}}, 2015, pp. 3460--3469.

\bibitem{kraus2016classifying}
O.~Z. Kraus, L.~J. Ba, and B.~J. Frey, ``Classifying and segmenting microscopy
  images with deep multiple instance learning,'' \emph{Bioinformatics},
  vol.~32, no.~12, pp. 52--59, 2016.

\bibitem{papandreou2015modeling}
X.~Wang, Y.~Yan, P.~Tang, X.~Bai, and W.~Liu, ``Revisiting multiple instance
  neural networks,'' \emph{Pattern Recognition}, vol.~74, pp. 15--24, 2018.

\bibitem{liu2018deep}
X.~Liu, L.~Jiao, J.~Zhao, J.~Zhao, D.~Zhang, F.~Liu, S.~Yang, and X.~Tang,
  ``{Deep Multiple Instance Learning-Based Spatial-Spectral Classification for
  {PAN} and {MS} Imagery},'' \emph{{IEEE} Trans. Geoscience and Remote
  Sensing}, vol.~56, no.~1, pp. 461--473, 2018.

\bibitem{ilse2018attention}
M.~Ilse, J.~M. Tomczak, and M.~Welling, ``{Attention-based Deep Multiple
  Instance Learning},'' in \emph{International Conference on Machine Learning,
  {ICML}}, 2018, pp. 2132--2141.

\bibitem{DBLP:conf/icml/DauphinFAG17}
Y.~N. Dauphin, A.~Fan, M.~Auli, and D.~Grangier, ``{Language Modeling with
  Gated Convolutional Networks},'' in \emph{International Conference on Machine
  Learning, {ICML}}, 2017, pp. 933--941.

\bibitem{DBLP:conf/icml/LiuWYY16}
W.~Liu, Y.~Wen, Z.~Yu, and M.~Yang, ``{Large-Margin Softmax Loss for
  Convolutional Neural Networks},'' in \emph{International Conference on
  Machine Learning, {ICML}}, 2016, pp. 507--516.

\bibitem{DBLP:conf/cvpr/LiuWYLRS17}
W.~Liu, Y.~Wen, Z.~Yu, M.~Li, B.~Raj, and L.~Song, ``{SphereFace: Deep
  Hypersphere Embedding for Face Recognition},'' in \emph{{IEEE} Conference on
  Computer Vision and Pattern Recognition, {CVPR}}, 2017, pp. 6738--6746.

\bibitem{DBLP:conf/cvpr/MignonJ12}
A.~Mignon and F.~Jurie, ``{{PCCA:} {A} new approach for distance learning from
  sparse pairwise constraints},'' in \emph{{IEEE} Conference on Computer Vision
  and Pattern Recognition, {CVPR}}, 2012, pp. 2666--2672.

\bibitem{godfrey1992switchboard}
J.~J. Godfrey, E.~C. Holliman, and J.~McDaniel, ``{SWITCHBOARD: Telephone
  speech corpus for research and development},'' in \emph{{IEEE} International
  Conference on Acoustics, Speech and Signal Processing, {ICASSP}}, 1992, pp.
  517--520.

\bibitem{xiong2017microsoft}
W.~Xiong, J.~Droppo, X.~Huang, F.~Seide, M.~Seltzer, A.~Stolcke, D.~Yu, and
  G.~Zweig, ``{The Microsoft 2016 conversational speech recognition system},''
  in \emph{{IEEE} International Conference on Acoustics, Speech and Signal
  Processing, {ICASSP}}, 2017, pp. 5255--5259.

\bibitem{vesely2013sequence}
K.~Vesel{\`y}, A.~Ghoshal, L.~Burget, and D.~Povey, ``Sequence-discriminative
  training of deep neural networks.'' in \emph{{INTERSPEECH}}, 2013, pp.
  2345--2349.

\bibitem{povey2016purely}
D.~Povey, V.~Peddinti, D.~Galvez, P.~Ghahremani, V.~Manohar, X.~Na, Y.~Wang,
  and S.~Khudanpur, ``{Purely Sequence-Trained Neural Networks for ASR Based on
  Lattice-Free MMI.}'' in \emph{{INTERSPEECH}}, 2016, pp. 2751--2755.

\bibitem{medennikov2016improving}
W.~Hartmann, R.~Hsiao, T.~Ng, J.~Ma, F.~Keith, and M.-H. Siu, ``{Improved
  Single System Conversational Telephone Speech Recognition with VGG Bottleneck
  Features},'' in \emph{{INTERSPEECH}}, 2017, pp. 112--116.

\bibitem{chan2016listen}
W.~Chan, N.~Jaitly, Q.~V. Le, and O.~Vinyals, ``{Listen, attend and spell: {A}
  neural network for large vocabulary conversational speech recognition},'' in
  \emph{{IEEE} International Conference on Acoustics, Speech and Signal
  Processing, {ICASSP}}, 2016, pp. 4960--4964.

\bibitem{DBLP:conf/interspeech/WatanabeHKHNUSH18}
S.~Watanabe, T.~Hori, S.~Karita, T.~Hayashi, J.~Nishitoba, Y.~Unno, N.~E.~Y.
  Soplin, J.~Heymann, M.~Wiesner, N.~Chen, A.~Renduchintala, and T.~Ochiai,
  ``{ESPnet: End-to-End Speech Processing Toolkit},'' in \emph{{INTERSPEECH}},
  2018, pp. 2207--2211.

\bibitem{DBLP:journals/corr/KingmaB14}
D.~P. Kingma and J.~Ba, ``{Adam: {A} Method for Stochastic Optimization},'' in
  \emph{International Conference on Learning Representations, {ICLR}}, 2015.

\bibitem{DBLP:conf/emnlp/Kim14}
Y.~Kim, ``Convolutional neural networks for sentence classification,'' in
  \emph{Conference on Empirical Methods in Natural Language Processing,
  {EMNLP}}, 2014, pp. 1746--1751.

\end{thebibliography}
\end{document}